\documentclass [12pt,a4paper]{article}
\usepackage{amssymb}
\newcommand{\<}{\langle}
\renewcommand{\>}{\rangle}

\def\bbbr{{\mathbb R}}
\def\bbbc{{\mathbb C}}

\def\pont{\,\cdot\,}
\def\Tr{\mbox{Tr}\,}

\def\supp{\mbox{supp}\,}
\def\Diag{\mbox{Diag}\,}
\def\tr{\mbox{Tr}\,}
\def\ti{\mbox{Tr}}
\def\tti{\mbox{tr}}
\def\qed{{\hfill $\square$}\medskip}

\def\bbbz{{ \mathbb Z}}

\def\iH{{\cal H}}
\def\iA{{\cal A}}
\def\iE{{\cal E}}
\def\iK{{\cal K}}
\def\Prob{\hbox{Prob}}

\newtheorem{thm}{Theorem}
\newtheorem{lemma}{Lemma}
\newtheorem{corollary}{Corollary}
\newtheorem{example}{Example}
\begin{document}
\ \vskip 1cm
\centerline{\LARGE Monotonicity}
\medskip
\centerline{\LARGE of quantum relative entropy revisited}
\bigskip
\medskip
\centerline{\bf This paper is dedicated to Elliott Lieb and Huzihiro
Araki}
\centerline{\bf on the occasion of their 70th birthday}
\bigskip
\medskip
\centerline{\large D\'enes Petz\footnote{E-mail: petz{@}math.bme.hu. The
work was supported by the Hungarian OTKA T032662.}}
\bigskip

\bigskip
\centerline{Department for Mathematical Analysis}
\centerline{Budapest University of Technology and Economics}
\centerline{ H-1521 Budapest XI., Hungary}
\bigskip\bigskip
\noindent
\begin{quote}
Monotonicity under coarse-graining is a crucial property of the
quantum relative entropy. The aim of this paper is to investigate 
the condition of equality in the monotonicity theorem and in its 
consequences as the strong sub-additivity of von Neumann entropy, the 
Golden-Thompson trace inequality and the monotonicity of the Holevo 
quantitity. The relation to quantum Markov states is briefly indicated.
\end{quote}

\begin{quote}
\end{quote}
\section{Introduction}
Quantum relative entropy was introduced by Umegaki \cite{Um} as a 
formal generalization of the Kullback-Leibler information (in the
setting of finite von Neumann algebras). Its real importance was 
understood much later and the monograph \cite{OP} already deduced
most information quantities from the relative entropy.

One of the fundamental results of quantum information theory is the 
monotonicity of relative entropy under completely positive mappings.
After the discussion of some particular cases by Araki \cite{Araki1976} 
and by Lindblad \cite{Lindblad},
this result was proven by Uhlmann \cite{Uh} in full generality and
nowadays it is 
referred as {\it Uhlmann's theorem}. The strong sub-additivity property of
entropy can be obtained easily from Uhlmann's theorem (see \cite{OP} 
about this point and as a general reference as well) and Ruskai 
discussed the relation of several basic entropy inequalities in 
details \cite{MBR0, MBR1}. The aim of this paper is to investigate 
the condition
of equality in the monotonicity theorem and in its consequences.
The motivation to do this comes from the needs of quantum information
theory developed in the setting of matrix algebras in the last ten 
years \cite{N-Ch}, on the other hand work \cite{MBR1} has given also some 
stimulation. 

The paper is written entirely in a finite dimensional
setting but some remarks are made about the possible more general
scenario.

\section{Uhlmann's theorem}

Let $\iH$ be a finite dimensional Hilbert space and $D_i$ be statistical
operators on $\iH$ ($i=1,2$). Their {\it relative entropy} is defined as
\begin{equation}
 S(D_1,D_2)=\cases{\Tr D_1(\log D_1 -\log D_2)\quad &  if  $\supp D_1
\subset \supp D_2$, \cr +\infty  & otherwise .}
\end{equation}
If $\lambda>0$ is the smallest eigenvalue of $D_2$, then $S(D_1,D_2)$ is
always finite and $S(D_1,D_2)\le \log n -\log \lambda$, where $n$ is the 
dimension of $\iH$. 

Let $\iK$ be another finite dimensional Hilbert
space. We call a linear mapping $T: B(\iH)\to B(\iK)$ {\it coarse graining}
if $T$ is trace preserving and $2$-positive, that is
\begin{equation}
\left[ \begin{array}{cc}
T(A)&T(B)\\T(C)&T(D)\end{array} \right] \ge 0 \hbox{\ if }
\left[ \begin{array}{cc}
A& B\\ C& D \end{array} \right] \ge 0\,. 
\end{equation}
Such a $T$ sends a statistical operator to statistical operator and
satisfies the Schwarz inequality $T(a^*a)\ge T(a)T(a)^*$. The concept
of coarse graining is the quantum version of the Markovian mapping in
probability theory. All the important examples are actually completely
positive. We work in this more general framework because the proofs
require only the Schwarz inequality. 

$B(\iH)$ and $B(\iK$ are Hilbert spaces with respect to the
Hilbert-Schmidt inner product and the adjoint of $T: B(\iH)\to B(\iK)$
is defined:
$$
\tr A \,T(B)= \tr T^*(A)\, B \qquad (A\in B(\iK), B \in B(\iH)).
$$
The adjoint of a coarse graining $T$ is 2-positive again and
$T^*(I)=I$. It follows that $T^*$ satisfies  the Schwarz inequality as
well.

The following result is known as Uhlmann's theorem.

\begin{thm} (\cite{Uh, OP}) For a coarse graining $T:B(\iH)\to B(\iK)$ 
the monotonicity
$$
S(D_1,D_2) \ge S(T(D_1),T(D_2))
$$
holds. \label{T:U}
\end{thm}

It should be noted that relative entropy was defined in the setting of
von Neumann algebras first by Umegaki \cite{Um} and extended by Araki 
\cite{Araki1976}. Uhlmann's monotonicity result is more general 
than the above statement.
 
To the best knowledge of the author, it is not known if the
monotonicity theorem holds without the hypothesis of 2-positivity.

\section{The proof of Uhlmann's theorem and its analysis}

The simplest way to analyse the equality in the monotonicity theorem is
to have a close look at the proof of the inequality. Therefore we
present a proof which is based on the {\it relative modular operator method}.
The concept of relative modular operator was developed by Araki in
the modular theory of operator algebras \cite{ArakiMasuda}, however it 
could be used
very well in finite dimensional settings. For example, {\it Lieb's concavity
theorem} gets a natural proof by this method \cite{petz1986}.

Let $D_1$ and $D_2$ be density matrices acting on the Hilbert space
$\iH$ and assume that they are invertible. On the Hilbert space $B(\iH)$
one can define an operator $\Delta$ as
$$
\Delta a = D_2 a D_1^{-1} \qquad (a \in B(\iH)).
$$ 
This is the so-called relative modular operator and it is the product
of two commuting positive operators: $\Delta=LR$, where
$$
La= D_2a\quad {\rm and}\quad Ra=a D_1^{-1}  \qquad (a \in B(\iH)).
$$
Since $\log \Delta= \log L +\log R$, we have
$$
{S}(D_1, D_2) = \<D^{1/2}_1, (\log D_1 -\log D_2) D^{1/2}_1\>=
- \<D^{1/2}_1, (\log \Delta) D^{1/2}_1\>. 
$$
The relative entropy ${S}(D_1, D_2)$ is expressed by the quadratic form
of the logarithm of the relative modular operator. This is the
fundamental formula what we use (and actually this is nothing else but
Araki's definition of the relative entropy in a general von Neumann
algebra \cite{Araki1976}). 

Let $T$ be a coarse graining as in Theorem \ref{T:U}.
We assume that $D_1$ and $T(D_1)$ are invertible matrices and set
$$
\Delta a = D_2 a D_1^{-1} \quad (a \in B(\iH))\quad\hbox{and}\quad
\Delta_0 x = T(D_2) x T(D_1)^{-1} \quad (x \in B(\iK)).
$$
$\Delta$ and $\Delta_0$ are operators on the spaces $B(\iH)$ and 
$B(\iK)$, respectively. Both become a Hilbert space with the
Hilbert-Schmidt inner product. The relative entropies in the theorem
are expressed by the resolvent of relative modular operators:
\begin{eqnarray*}
{S}(D_1, D_2) &=& -\<D^{1/2}_1, (\log \Delta) D^{1/2}_1\> \\
&=& \int^{\infty}_{0} \<D^{1/2}_1, (\Delta + t)^{-1} D^{1/2}_1\>
-(1 + t)^{-1}\,dt \\
{S}(T(D_1), T(D_2)) &=& -\<T(D_1)^{1/2}, (\log \Delta_0) T(D_1)^{1/2}\> \\
&=& \int^{\infty}_{0} \<T(D_1)^{1/2}, (\Delta_0 + t)^{-1} 
T(D_1)^{1/2}\>-(1 + t)^{-1} \,dt,  
\end{eqnarray*}
where the identity 
$$
\log x= \int^{\infty}_{0} (1 + t)^{-1}-(x+t)^{-1}\,dt
$$
is used. The operator 
\begin{equation}
VxT(D_1)^{1/2}=T^*(x)D_1^{1/2} 
\end{equation}
is a contraction:
$$
\Vert T^*(x)D_1^{1/2}\Vert^2 = \Tr D_1 T^*(x^*)T^*(x)\le  
\Tr D_1 T^*(x^* x)=\Tr T(D_1) x^* x=\Vert xT(D_1)^{1/2}\Vert^2
$$
since the Schwarz inequality is applicable to $T^*$. A similar simple 
computation gives that
\begin{equation}
V^*\Delta V \le \Delta_0\,.
\end{equation}
The function $y \mapsto (y+t)^{-1}$ is operator monotone (decreasing) and
operator convex, hence
\begin{equation}\label{E:op}
(\Delta_0+t)^{-1}\le (V^* \Delta V +t)^{-1}\le V^*(\Delta +t)^{-1}V
\end{equation}
(see \cite{HanPed}). Since $VT(D_1)^{1/2}= D_1^{1/2}$,  this implies
\begin{equation}
\<D^{1/2}_1, (\Delta + t)^{-1} D^{1/2}_1\> \ge 
\<T(D_1)^{1/2}, (\Delta_0 + t)^{-1} T(D_1)^{1/2}\,. 
\end{equation}
By integrating this inequality we have the monotonicity theorem
from the above integral formulas.

Now we are in the position to analyse the case of equality. If
$$
S(D_1,D_2)= S(T(D_1),T(D_2)),
$$
then 
\begin{equation}
\<T(D_1)^{1/2}, V^*(\Delta + t)^{-1}V T(D_1)^{1/2}\> = 
\<T(D_1)^{1/2}, (\Delta_0 + t)^{-1} T(D_1)^{1/2}\>\,. 
\end{equation}
for all $t>0$. This equality together with the operator inequality
(\ref{E:op}) gives
\begin{equation}
V^*(\Delta + t)^{-1} D_1^{1/2}=(\Delta_0 + t)^{-1} T(D_1)^{1/2}
\end{equation}
for all $t>0$. Differentiating by $t$ we have
\begin{equation}
V^*(\Delta + t)^{-2} D_1^{1/2}=(\Delta_0 + t)^{-2} T(D_1)^{1/2}
\end{equation}
and we infer
\begin{eqnarray*}
\Vert V^*(\Delta + t)^{-1} D_1^{1/2}\Vert^2
&=&\<(\Delta_0 + t)^{-2} T(D_1)^{1/2}, T(D_1)^{1/2}\>\\ 
&=& \<V^*(\Delta + t)^{-2} D_1^{1/2},T(D_1)^{1/2}\>\\
&=& \Vert (\Delta + t)^{-1} D_1^{1/2}\Vert^2
\end{eqnarray*}
When $\Vert V^*\xi\Vert = \Vert \xi \Vert$ holds for a contraction $V$,
it follows that $VV^*\xi=\xi$. In the light of this remark we arrive at
the condition
$$
VV^*(\Delta + t)^{-1} D_1^{1/2}=(\Delta + t)^{-1} D_1^{1/2}
$$
and
\begin{eqnarray*}
V(\Delta_0 + t)^{-1} T(D_1)^{1/2}
& =& VV^*(\Delta + t)^{-1} D_1^{1/2}\\
& = & (\Delta + t)^{-1} D_1^{1/2}
\end{eqnarray*}
By Stone-Weierstrass approximation we have
\begin{equation}
V f(\Delta_0) T(D_1)^{1/2}=f(\Delta) D_1^{1/2}
\end{equation}
for continuous functions. In particular for $f(x)=x^{i t}$ we
have
\begin{equation}\label{E:ns}
T^*\big(T(D_2)^{it}T(D_1)^{-it}\big)= D_2^{it} D_1^{-it}\,. 
\end{equation}
This condition is necessary and sufficient for the equality. 

\begin{thm} \label{masodik}
Let $T:B(\iH)\to B(\iK)$ be a $2$-positive trace preserving 
mapping and let $D_1, D_2 \in B(\iH), T(D_1), T(D_2)\in B(\iK)$ be
invertible density matrices. Then the equality 
$
S(D_1,D_2) = S(T(D_1),T(D_2))
$
holds if and only if the following equivalent conditions are satisfied:
\begin{itemize}
\item[(1)] $T^*\big(T(D_1)^{it}T(D_2)^{-it}\big)= D_1^{it} D_2^{-it}$
for all real $t$.
\item[(2)] $T^*(\log T(D_1)-\log T(D_2))=\log D_1 -\log D_2 $.
\end{itemize}
\end{thm}
  
The equality implies (\ref{E:ns}) which is equivalent to (1) in the Theorem. 
Differentiating (1) at $t=0$
we have the second condition which obviously applies the equalities
of the relative entropies. \qed

The above proof follows the lines of \cite{petz1988}. The original paper is 
in the setting of arbitrary von Neumann algebras and hence slightly more
technical (due to the unbounded feature of the relative modular
operators). Condition (2) of Theoem \ref{masodik} appears also in the
paper \cite{MBR1} in which different methods are used.

Next we recall a property of 2-positive mappings. When $T$ is assumed to be
2-positive, the set
$$
\iA_{T}:=\{X\in B(\iH): T(X^*X)=T(X)T(X^*) \hbox{\ and\ }
T(X^*X)=T(X^*)T(X)\}.
$$
is a *-sub-algebra of $B(\iH)$ and
\begin{equation}
T(XY)=T(X)T(Y)\quad \hbox{for\ all\ } X\in \iA_T 
\hbox{\ and\ } Y\in B(\iH).
\end{equation}

\begin{corollary} Let $T:B(\iH)\to B(\iK)$ be a $2$-positive trace preserving 
mapping and let $D_1, D_2 \in B(\iH), T(D_1), T(D_2)\in B(\iK)$ be
invertible density matrices. Assume that $T(D_1)$ and $T(D_2)$
commute. Then the equality 
$
S(D_1,D_2) = S(T(D_1),T(D_2))
$
implies that $D_1$ and $D_2$ commute.
\end{corollary}

Under the hypothesis $u_t:=T(D_1)^{it}T(D_2)^{-it}$ and $w_t:=
D_1^{it} D_2^{-it}$ are unitaries. Since $T^*$ is unital $u_t \in
\iA_{T^*}$ for every $t \in \bbbr$. We have
$$
w_{t+s}=T^*(u_{t+s})=T^*(u_{t}u_s)=T^*(u_t)T^*(u_s)=w_t w_s
$$
which shows that $w_t$ and $w_s$ commute and so do $D_1$ and $D_2$.\qed

\section{Consequences and related inequalities}

{\bf 3.1. The Golden-Thompson inequality.} The Golden-Thompson
inequality tells that
$$
\Tr e^{A+B} \le \Tr e^A e^B
$$
holds for self-adjoint matrices $A$ and $B$. It was shown in \cite{petz1988b}
that this inequality can be reformulated as a particular case of
monotonicity when $e^A/\tr e^A$ is considered as a density matrix and
$e^{A+B}/\Tr e^{A+B}$ is the so-called perturbation by $B$. Corollary 
5 of the original paper is formulated in the context of von Neumann 
algebras but the argument was adapted to the finite dimensional case 
in \cite{petz1994}, see also p. 128 in \cite{OP}. The equality holds in
the Golden-Thompson inequality if and only if $AB=BA$.

One of the possible extensions of the Golden-Thompson inequality is the
statement that the function
\begin{equation}\label{E:Fried}
p \mapsto \Tr (e^{pB/2}e^{pA}e^{pB/2})^{1/p} 
\end{equation}
is increasing for $p>0$. The limit at $p=0$ is $\Tr e^{A+B}$
\cite{Araki1990}. It was proved by Friedland and So that the function
(\ref{E:Fried}) is strictly monotone or constant \cite{Friedland}. 
The latter case corresponds to the commutativity of $A$ and $B$.

{\bf 3.2. A posteriori relative entropy.}
Let $E_j$ ($1\le j \le m$) be a partition of unity in $B(\iH)_+$, that is
$\sum_j E_j=I$. (The operators $E_j$ could describe a measurement giving
finitely many possible outcomes.) Any density matrix $D_i \in B(\iH)$
determines a probability distribution
$$
\mu_i=(\Tr D_i E_1,\Tr D_iE_2, \dots, \Tr D_iE_m).
$$
It follows from Uhlmann's theorem that
\begin{equation}\label{E:egy}
S(\mu_1, \mu_2) \le S(D_1,D_2)\, .
\end{equation}

We give an example that the equality in (\ref{E:egy}) may appear
non-trivially.

\begin{example} Let $D_2=\Diag(1/3,1/3,1/3)$, $D_1=\Diag(1-2\mu, \mu,\mu)$,
$E_1=\Diag(1,0,0)$ and
$$
E_2=\left[\begin{array}{ccc} 0 & 0 & 0 \\ 0 & x & z\\ 0 & \bar{z}&1-x
\end{array} \right],\qquad
E_3=\left[\begin{array}{ccc} 0 & 0 & 0 \\ 0 & 1-x & -z\\ 0 & -\bar{z}& x
\end{array} \right]
$$
When $0 < \mu <1/2$, $0<x<1$ and for the complex $z$ the modulus of $z$
is small enough we have a partition of unity and $S(\mu_1, \mu_2) =
S(D_1,D_2)$ holds.
\end{example}

First we prove a lemma.

\begin{lemma} If $D_2$ is an invertible density then the
equality in (\ref{E:egy}) implies that $D_2$ commutes with $D_1, E_1,
E_2,\dots, E_m$.
\end{lemma}

The linear operator $T$ associates a diagonal matrix 
$$
\Diag(\Tr D E_1,\Tr D E_2,  \dots, \Tr DE_m))
$$ 
to the density $D$ acting on $\iH$
and under the hypothesis (\ref{E:ns}) is at our disposal. We have
$$
\<D_2^{1/2}, T^*\big(T(D_1)^{it}T(D_2)^{-it}\big)D_2^{1/2}\>=
\<D_2^{1/2}, D_1^{it} D_2^{-it}D_2^{1/2}\>\,.
$$
Actually we
benefit from the analytic continuation and we put $-i/2$ in place of $t$.
Hence
\begin{equation}
\sum_{j=1}^m (\Tr E_j D_1)^{1/2}(\Tr E_j D_1)^{1/2}=\Tr D_1^{1/2} D_2^{1/2}\,.
\end{equation}

The Schwarz inequality tells us that
\begin{eqnarray*}
\Tr D_1^{1/2} D_2^{1/2}&=&\< D_1^{1/2}, D_2^{1/2}\>=\sum_{j=1}^m
\< D_1^{1/2}E_j^{1/2}, D_2^{1/2}E_j^{1/2}\>\\
&\le&
\sum_{j=1}^m 
\sqrt{\< D_1^{1/2}E_j^{1/2}, D_1^{1/2}E_j^{1/2}\>}
\sqrt{\< D_2^{1/2}E_j^{1/2}, D_2^{1/2}E_j^{1/2}\>} \\
&=&
\sum_{j=1}^m (\Tr E_j D_1)^{1/2}(\Tr E_j D_2)^{1/2}\,.
\end{eqnarray*}
The condition for equality in the Schwarz inequality is well-known:
There are some complex numbers $\lambda_j \in \bbbc$ such that
\begin{equation}\label{E:def}
D_1^{1/2}E_j^{1/2}= \lambda_j  D_2^{1/2}E_j^{1/2}\,. 
\end{equation}
(Since both sides have positive trace, $\lambda_j$ are actually positive. )
The operators $E_j$ and $E_j^{1/2}$ have the same range, therefore
\begin{equation}\label{E:2}
D_1^{1/2}E_j = \lambda_j  D_2^{1/2}E_j\,. 
\end{equation}
Summing over $j$ we obtain
$$
D_2^{-1/2}D_1^{1/2} = \sum_{j=1}^m \lambda_j E_j\,.
$$
Here the right hand side is self-adjoint, so $D_2^{-1/2}D_1^{1/2} =
D_1^{1/2}D_2^{-1/2} $ and $D_1D_2=D_2D_1$. Now it follows from
(\ref{E:def}) that $E_j$ commutes with $D_2$. \qed

Next we analyse the equality in (\ref{E:egy}).
If $D_2$ is invertible, then the previous lemma tells us that $D_1$ and
$D_2$ are diagonal in an appropriate basis. In this case $S(\mu_1,
\mu_2)$ is determined by the diagonal elements of the matrices $E_j$.
Let $\iE(A)$ denote the diagonal matrix whose diagonal coincides with
that of $A$. If $E_j$ is a partition of unity, then so is $\iE(E_j)$.
However, given a partition of unity $F_j$ of diagonal matrices, there
could be many choice of a partition of unity $E_j$ such that $\iE(E_j)
=F_j$, in general. In the moment we do not want to deal with this
ambiguity, and we
assume that we have a basis $e_1, e_2,\dots,e_n$ consisting of common
eigenvectors of the operators $D_1,D_2,\iE(E_1),\iE(E_2),\dots,\iE(E_n)$:
$$
D_i e_k= v^i_ke_k \quad\hbox{and}\quad  
\iE(E_j) e_k= w_{kj} e_k \quad(i=1,2,\, j=1,2,\dots,m,\,k=1,2, \dots, n).
$$
The matrix $[w_{kj}]_{kj}$ is (raw) stochastic and condition
(\ref{E:2}) gives
$$
\frac{v^1_k}{v^2_k}w_{kj}=(\lambda_j)^2 w_{kj}
$$
This means that $w_{kj}\ne 0$ implies that ${v^1_k}/{v^2_k}$ does not
depend on $k$. In other words, $D_1D_2^{-1}$ is constant on the support
of any $E_j$.

Let $j$ be equivalent with $k$, if the support of $\iE(E_j)$ intersects
the support of $\iE(E_k)$. We denote by $[j]$ the equivalence class of $j$
and let $J$ be the set of equivalence classes.
$$
P_{[j]}:=\sum_{k\in[j]} \iE(E_k)
$$
must be a projection and $\{P_{[j]}: [j]\in J\}$
is a partition of unity. We deduced above that
$$
D_1D_2^{-1}P_{[j]}=\lambda_j P_{[j]}
$$
One cannot say more about the condition for equality. All these
extracted conditions hold in the above example and $\iE(E_k)$'s do not
determine $E_k$'s, see the freedom for the variable $z$ in the example.

We can summarise our analysis as follows. The case of equality in 
(\ref{E:egy}) implies some commutation relation and the whole problem
is reduced to the commutative case. It is not necessary that the
positive-operator-valued measure $E_j$ should have projection values.

{\bf 3.3. The Holevo bound.} Let $E_j$ ($1\le j \le m$) be a partition 
of unity in $B(\iK)_+$, $\sum_j E_j=I$. We assume that the density
matrix $D\in B(\iH)$ is in the form of a convex combination $D=\sum_i 
p_iD_i$ of other densities $D_i$. Given a coarse graining $T:B(\iH)\to 
B(\iK)$ we can say that
our signal $i$ appears with probability $p_i$, it is encoded by the 
density matrix $D_i$, after transmission the density $T(D_i)$ appears
in the output and the receiver decides that the signal $j$ was sent
with the probability $\tr T(D_i)E_j$. This is the standard scheme of 
quantum information transmission. Any density matrix $D_i \in B(\iH)$
determines a probability distribution
$$
\mu_i=(\Tr T(D_i) E_1,\Tr T(D_i) E_2, \dots, \Tr T(D_i)E_m).
$$
on the output. The inequality
\begin{equation}\label{E:Hb}
S(\mu)-\sum_i p_i S(\mu_i) \le S(D)-\sum_i p_i S(D_i)
\end{equation}
(where $\mu:=\sum_i p_i \mu_i$ and $D:=\sum_i p_i D_i$) is the
so-called {\it Holevo bound} for the amount of information passing through the 
communication channel. Note that the Holevo bound appeared before the
use of quantum relative entropy and the first proof was more complicated.

$\mu_i$ is a coarse-graining of $T(D_i)$, therefore inequality
(\ref{E:Hb}) is of the form
$$
\sum_i p_iS(R(D_i), R(D)) \le \sum_i p_i S(D_i,D).
$$
On the one hand, this form shows that the bound (\ref{E:Hb}) is a 
consequence of the monotonicity, on the other hand, we can make an
analysis of the equality. Since the states $D_i$ are the codes of the
messages to be transmitted, it would be too much to assume that all of
them are invertible. However, we may assume that $D$ and $T(D)$ are 
invertible. Under this hypothesis Lemma 1 applies and tells us that
the equality in (\ref{E:Hb}) implies that all the operators $T(D),
T(D_i)$ and $E_j$ commute. 

{\bf 3.4. $\alpha$-entropies.} The $\alpha$-divergence of the densities
$D_1$ and $D_2$ is
\begin{equation}
S_{\alpha} (D_1 , D_2) = \frac{4}{1- \alpha^2}{\Tr}\:
(D_1 - D_1^{\frac{1+\alpha}{2}} D_2^{\frac{1-\alpha}{2}}), 
\end{equation}
which is essentially
$$
\<D^{1/2}_2, \Delta^{\frac{1+\alpha}{2}} D^{1/2}_2\> 
$$
up to constants in the notation of Sect. 2. The proof of the
monotonicity works for this more general quantity with a small
alteration. What we need is
$$
\<D^{1/2}_2, \Delta^\beta D^{1/2}_2\> =\frac{\sin \pi \beta}{\pi}
\int^{\infty}_{0} -t^\beta\<D^{1/2}_2, (\Delta + t)^{-1} D^{1/2}_2\>+
t^{\beta-1} \,dt 
$$
for $0 < \beta <1$. Therefore for $0< \alpha <2$ the proof of the above
Theorem 2 goes through for the $\alpha$-entropies. The monotonicity
holds for the $\alpha$-entropies, moreover (1) and (2) from Theorem
\ref{masodik} are necessary and sufficient for the equality.
 
The role of the $\alpha$-entropies is smaller than that of the relative
entropy but they are used for approximation of the relative entropy and
for some other purposes (see \cite{hape}, for example).

\section{Strong subadditivity of entropy and the Markov property}
The {\it strong subadditivity} is a crucial property of the von Neumann entropy
it follows easily from the monotonicity of the relative entropy. (The
first proof of this property of entropy was given by Lieb and Ruskai 
\cite{SSA} before the Uhlmann's monotonicity theorem.)
The strong subadditivity property is related to the composition of three 
different systems. It is used, for example,  in the analysis of the
translation invariant states of quantum lattice systems: The proof of
the existence of the global entropy density functional is based on the
subadditivity and a monotonicity property  of local entropies is 
obtained by the strong subadditivity \cite{petz1996}.

Consider three Hilbert spaces, $\iH_j$, $j=1,2,3$ and a
statistical operator $D_{123}$ on the tensor product $\iH_1\otimes 
\iH_2\otimes \iH_3$. This statistical operator has marginals on all
subproducts, let $D_{12}$, $D_{2}$ and $D_{23}$ be the marginals on 
$\iH_1\otimes \iH_2$, $\iH_2$ and $\iH_2\otimes \iH_3$, respectively.
(For example, $D_{12}$ is determined by the requirement $\tr D_{123}
(A_{12} \otimes I_3)=\tr D_{12}A_{12}$ for every operator $A_{12}$ acting
on $\iH_1\otimes  \iH_2$; $D_{2}$ and $D_{23}$ are similarly defined.)
The strong subadditivity asserts the following:
\begin{equation}\label{E:ssa}
S(D_{123})+S(D_2)\le
S(D_{12})+S(D_{23})
\end{equation}

In order to prove the strong subadditivity, one can start with the 
identities
\begin{eqnarray*}
S(D_{123}, \tti_{123})&=&S(D_{12},\tti _{12})+S(D_{123},
D_{12}\otimes \tti _3)
\\
S(D_2,\tti _2)+S(D_{23},D_2\otimes \tti _3)&=&S(D_{23},
\tti _{23}),
\end{eqnarray*}
where $\tti$ with a subscript denotes the density of the corresponding
tracial state, for example $\tti _{12}=I_{12}/\dim (\iH_1\otimes \iH_2)$. 
From these equalities we arrive at a new one,
\begin{eqnarray*}
S(D_{123}, \tti_{123})+ S(D_2,\tti _2) &=&
S(D_{12},\tti _{12})+ S(D_{23}, \tti _{23})\\ &\quad & +
S(D_{123}, D_{12}\otimes \tti _3)-S(D_{23},D_2\otimes \tti _3).
\end{eqnarray*}
If we know that
\begin{equation}\label{E:RSS}
S(D_{123},D_{12}\otimes \tti _3)\ge S(D_{23}, D_2\otimes
\tti _3)
\end{equation}
then the strong subadditivity (\ref{E:ssa}) follows. Set a linear 
transformation $B(\iH_1\otimes \iH_2\otimes \iH_3)\to B(\iH_2\otimes
\iH_3)$ as follows:
\begin{equation}\label{E:T}
T (A\otimes B\otimes C):= B\otimes C (\Tr A ) 
\end{equation}
T is completely positive and trace preserving. On the other hand,
$T(D_{123})=D_{23}$ and $T(D_{12}\otimes \tti _3)=D_{2}\otimes \tti _3$.
Hence the monotonicity theorem gives (\ref{E:RSS}). 

This proof is very transparent and makes the equality case visible.
The equality in the strong subadditivity holds if and only if we have
equality in (\ref{E:RSS}). Note that $T$ is the partial trace over the
third system and
\begin{equation}\label{E:Tcsil}
T^*(B\otimes C)=I\otimes B \otimes C.
\end{equation}

\begin{thm}\label{equality}
Assume that $D_{123}$ is invertible. The equality holds in the strong 
subadditivity (\ref{E:ssa}) if and only if the following equivalent 
conditions hold:
\begin{itemize}
\item[(1)] $D_{123}^{it} D_{12}^{-it}= D_{23}^{it} D_2^{-it}$
for all real $t$.
\item[(2)] $\log D_{123}-\log D_{12}=\log D_{23} -\log D_2 $.
\end{itemize}
\end{thm} 

Note that both condition (1) and (2) contain implicitely tensor
products, all operators should be viewed in the three-fold-product.
Theorem 2 applies due to (\ref{E:Tcsil}) and this is the proof.  \qed

It is not obvious the meaning of conditions (1) and (2) in
Theorem \ref{equality}. The easy choice is
$$
\log D_{12}=H_1+H_2+H_{12},\,\, \log D_{23}=H_2+H_3+H_{23},\,\, \log D_2=H_2
$$
for a commutative family of self-adjoint operators $H_1,H_2,H_3,H_{12},
H_{23}$ and to define $\log D_{123}$ by condition (2) itself. This
example lives in an abelian subalgebra of $\iH_1\otimes \iH_2\otimes 
\iH_3$ and a probabilistic representation can be given.

$D_{123}$ may be regarded as the joint probability distribution of some 
random variables $\xi_1, \xi_2$ and $\xi_3$. In this language we can
rewrite (1) in the form
\begin{equation}
\frac{\Prob (\xi_1= x_1, \xi_2= x_2, \xi_3= x_3 )}
{\Prob(\xi_1= x_1, \xi_2= x_2 )}= 
\frac{\Prob (\xi_2= x_2, \xi_3= x_3 )}
{\Prob(\xi_2= t_2 )}
\end{equation}
or in terms of conditional probabilities
\begin{equation}
\Prob (\xi_3= x_3 | \xi_1= x_1, \xi_2= x_2 )= 
\Prob (\xi_3= x_3 | \xi_2= x_2 )\,.
\end{equation}
In this form one recognizes the Markov property for the variables
$\xi_1, \xi_2$ and $\xi_3$; subscripts 1,2 and 3 stand for ``past'', 
``present'' and ``future''. It must be well-known that for classical
random variables the equality case in the strong subadditivity of the
entropy is equivalent to the Markov property.
The equality
\begin{equation}\label{E:inc}
S(D_{123})-S(D_{12}) = S(D_{23})-S(D_2)
\end{equation} 
means an equality of entropy increments. Concerning the Markov property, see
\cite{AF} or pp. 200--203 in \cite{OP}.

\begin{thm}
Assume that $D_{123}$ is invertible. The equality holds in the strong 
subadditivity (\ref{E:ssa}) if and only if there exists a completely
positive unital mapping $\gamma: B(\iH_1\otimes \iH_2\otimes \iH_3)\to 
B(\iH_2\otimes \iH_3)$ such that
\begin{itemize}
\item[(1)] $\tr(D_{123}\gamma(x))=\tr(D_{123}x)$  for all $x$.
\item[(2)] $\gamma|B(\iH_2)\equiv $ identity.
\end{itemize}
\end{thm} 

If $\gamma$ has properties (1) and (2), then $\gamma^*(D_{23})=D_{123}$
and $\gamma^*(D_2\otimes \ti_3)=D_{12}\otimes \ti_3$ for its dual and
we have equality in (\ref{E:RSS}).

To prove the converse let
\begin{equation}\label{E:E}
E (A\otimes B\otimes C):= B\otimes C (\Tr A /\dim \iH_1) 
\end{equation}
which is completely positive and unital. Set
\begin{equation}\label{E:gamma}
\gamma (\pont):= D_{23}^{-1/2}E(D_{123}^{1/2}\pont D_{123}^{1/2}) D_{23}^{-1/2}
\end{equation}
If the equality holds in the strong subadditivity, then property (1)
from Theorem 3 is at our disposal and it gives $\gamma(x)=x$ for $x \in
 B(\iH_2)$.\qed

In a probabilistic interpretation $E$ and $\gamma$ are conditional 
expectations. $E$ preserves the tracial state and it is a projection of
norm one. $\gamma$ leaves the state with density $D_{123}$ invariant,
however it is not a projection. (Accardi and Cecchini called this
$\gamma$ generalised conditional expectation, \cite{AC}.) 

It is interesting to construct translation invariant states on the
infinite tensor product of matrix algebras (that is, quantum spin chain
over $\bbbz$) such that condition (\ref{E:inc}) holds for all ordered
subsystems 1,2 and 3.

\end{document}